\documentclass[]{aastex631}

\begin{document}

\title{The discovery of blue-cored dwarf early-type galaxies in isolated environments}

\shorttitle{Discovering blue-cored dEs in isolated environments}

\author[0000-0002-0041-6490]{Soo-Chang Rey}
\affiliation{Department of Astronomy and Space Science, Chungnam National University, Daejeon 34134, Republic of Korea; screy@cnu.ac.kr}

\author[0000-0002-3738-885X]{Suk Kim}
\affiliation{Department of Astronomy and Space Science \& Research Institute of Natural Sciences, Chungnam National University, Daejeon 34134, Republic of Korea}

\author[0000-0003-0469-345X]{Jiwon Chung}
\affil{Korea Astronomy and Space Science Institute 776, Daedeokdae-ro, Yuseong-gu, Daejeon 34055, Republic of Korea}

\author[0000-0002-6261-1531]{Youngdae Lee}
\affil{Department of Astronomy and Space Science \& Research Institute of Natural Sciences, Chungnam National University, Daejeon 34134, Republic of Korea}




\begin{abstract}

{The presence of blue-cored dwarf early-type galaxies (dE(bc)s) in high-density environments supports the scenario of the transformation of infalling late-type galaxies into quiescent dwarf early-type galaxies by environmental effects. While low-density environments lacking environmental processes could not be relevant to the formation of dE(bc)s, 
we discovered a large sample of rare dE(bc)s in isolated environments at $z < 0.01$ using the NASA–Sloan Atlas catalog. Thirty-two isolated dE(bc)s were identified by visual inspection of the Sloan Digital Sky Survey images and $g-r$ color profiles. We found that (1) isolated dE(bc)s exhibit similar structural parameters to dE(bc)s in the Virgo cluster; (2) based on the ultraviolet$-r$ color-magnitude relation, color gradients, and optical emission lines of dE(bc)s, isolated dE(bc)s show more vigorous, centrally concentrated SF compared to their counterparts in the Virgo cluster; (3) at a given stellar mass, isolated dE(bc)s tend to have a larger fraction of gas mass than their Virgo counterparts. We discuss a scenario of episodic SF sustained by gas accretion, suggested by Sánchez Almeida et al., in which the star-bursting blue compact dwarf galaxy (BCD)$-$quiescent BCD (QBCD) cycle can be repeated during the Hubble time. We suggest that, in this cadence, isolated dE(bc)s might be QBCDs at pre- or post-BCD stages. 
Our results imply that dE(bc)s comprise a mixture of objects with two types of origins, nature or nurture, depending on their environment. }

\end{abstract}



\section{Introduction} \label{sec:intro}
Dwarf early-type galaxies (dEs), such as dwarf ellipticals and dwarf lenticulars, are the most numerous low-luminosity systems in galaxy clusters and groups \citep{Binggeli1988,Ferguson1994}. They show only a simple appearance of elliptical shape and are thought to be dominated by a homogeneous, old stellar population. However, unexpected complexity and diversity in their characteristics have recently become evident \citep[e.g.,][]{Lisker2006a,Lisker2006b,Paudel2010,Janz2012,Toloba2014,Ann2015,Michea2022}. 

A particularly intriguing subclass of dEs hosting substructures is that with a blue core, dubbed blue-cored dEs \citep[dE(bc)s;][]{Lisker2006a,Pak2014,Urich2017,Chung2019,Hamraz2019}. These are morphologically classified as dEs and the colors of their outer parts are comparable to those of normal, red dEs without blue cores. However, they exhibit central blue colors caused by young stellar populations formed from recent or ongoing central star formation (SF) \citep{Lisker2006a,Pak2014, Urich2017}. These galaxies have mostly been found in clusters and groups, indicating that environmental processes may play a part in their formation and evolution. 

A few different formation mechanisms have been proposed for dE(bc)s in high-density environments. The most promising mechanisms involve ram-pressure stripping, galaxy harassment, and galaxy merging, which are all responsible for transforming late-type galaxies into dEs (\citealt{Chung2019}, and references therein). However, all mechanisms appear to reproduce only part of the observational properties of dE(bc)s. 

In the absence of environmental effects, the formation of dE(bc)s in isolated environments could not be anticipated.
However, the discovery of a dE(bc) in isolation \citep[e.g., IC 225,][]{Gu2006} has challenged the formation scenarios of dE(bc)s through environmental effects in dense environments. For example, in contrast to the cluster environment, ram pressure stripping and galaxy harassment would not operate to form isolated dE(bc)s nor to regulate their subsequent evolution.

To investigate the mechanism responsible for forming dE(bc)s in isolation, a study of the properties of a large sample of isolated dE(bc)s is required. However, a systematic study for such objects has not been conducted yet. 
In this work, we present the discovery of a large sample of isolated dE(bc)s in the local universe. We discuss a possible origin of isolated dE(bc)s by comparing their stellar population properties with dE(bc)s in the Virgo cluster. Throughout this paper, we assume the cosmological parameters to be  $\Omega_{m} = 0.3$, $\Omega_{\Lambda} = 0.7$, and $h_{0} = 0.73$ \citep{Komatsu2011}.

\section{Sample Selection and Analysis} \label{sec:style}
For our analysis, photometric and spectroscopic parameters of galaxies are taken from the NASA-Sloan Atlas (NSA) catalog\footnote{http://nsatlas.org/}, which is mainly based on the Sloan Digital Sky Survey (SDSS) optical data with the addition of the Galaxy Evolution Explorer ultraviolet (UV) data. We selected 505 isolated galaxies using a cylindrical volume centered on each galaxy in the local universe of $cz$ $<$ 3,000 kms$^{-1}$ extracted from the NSA. We adopted the circle radius and depth of the cylindrical volume as  350 kpc and ±700 kms$^{-1}$, respectively. We defined an isolated galaxy as having no neighbors within the cylindrical volume  \citep[see][for details]{Chung2021}. 

To parameterize the environment surrounding each isolated galaxy, we computed the three-dimensional density contrast, i.e., the over-density defined by \citet{Gavazzi2010} as:

\begin{equation}
   \delta_{1,1000}= \frac{\rho-\langle{\rho}\rangle} {\langle{\rho}\rangle}
\end{equation}

, where $\rho$ is the local galaxy number density and $\langle$$\rho$$\rangle$  is the mean galaxy number density measured in the Coma/A1367 supercluster region (i.e., 0.05 galaxy $h^{-1}$ Mpc$^{-3}$). The $\rho$ around each isolated galaxy was computed within a cylinder of 1 $h^{-1}$ Mpc radius and 1000 kms$^{-1}$ half-length. We found that most of the isolated galaxies show values of $\delta$$_{1,1000}$ $<$ 0, indicating that they are located in an ultra-low-density environment \citep{Gavazzi2010}.

Among the isolated galaxies, we selected 336 dwarf galaxies by requesting $M_{r}$ $>$ $-$19 mag. Based on the reddening law of \citet{Cardelli1989}, Galactic extinction was corrected for the magnitudes of galaxies using the E($B-V$) value. The distances of galaxies were calculated from their redshifts provided by the NSA, which were estimated using a peculiar velocity model. We also used galaxies in the Extended Virgo Cluster Catalog \citep[EVCC;][]{Kim2014} for an adequate comparison with galaxies in a cluster environment. We adopted a distance modulus of the Virgo cluster of 31.1 mag \citep{Mei2007}.

Each of the four authors independently performed a visual inspection of the SDSS $g$, $r$, and $i$ bands combined color images of isolated dwarf galaxies, following the morphological classification scheme used for constructing the EVCC. The morphology of each galaxy was considered to be final if the classifications of three or more authors agree. Finally, we identified 35 bonafide isolated dEs that are characterized by spheroidal shapes with smooth surface brightness profiles and overall red colors. The remaining 301 dwarf galaxies consist of high- or low-surface brightness dwarf irregular galaxies.

We carried out the surface photometry of the isolated dEs in the $g$ and $r$ bands using the IRAF ELLIPSE task. We also fit the surface brightness profiles with a single S\'{e}rsic function, excluding data inside the inner region ($\sim$ 1.4 arcsec) to minimize the seeing effect. In Figure 1 (upper panels), we present examples of the S\'{e}rsic profile fitting to the observed $r$ band surface brightness profiles of galaxies.
The S\'{e}rsic indices of the majority of isolated dEs are in the range 0.9 $< n <$ 2.1, suggesting that they follow typical exponential profiles of cluster dEs \citep{Graham2003}. We estimated the global photometric parameters in the $r$ band, such as effective radius $R_{e}$ and mean effective surface brightness $<\mu>$$_{e}$ within $R_{e}$. We also derived these parameters of Virgo dwarf galaxies in the EVCC.

We derived the $g-r$ color gradient, defined by $d(g-r)/d$log($R$/$R_{e}$), using the linear least-square fit to the $g-r$ color profile.
To minimize the seeing effect and noisy data at larger radii, we used a fitting range from 1.4 arcsec to an effective radius. Out of 35  dEs, 32 galaxies (91 \%) were finally defined as dE(bc)s, which show positive color gradients. 
Most dE(bc)s exhibit color gradients of more than 0.2 mag dex$^{-1}$. The colors of central regions of isolated dE(bc)s are bluer than those of the rest of the galaxies in color images and color profiles (see insets and lower panels of Figure 1), which is quite compatible with the morphologies of dE(bc)s in high-density environments \citep{Lisker2006a,Pak2014}. The outer regions of all isolated dE(bc)s have remarkably red colors comparable to those of typical dEs without blue cores in the Virgo cluster (lower panels of Figure 1, grey-shaded areas). This indicates that the stellar populations of outer regions in dE(bc)s are similar to old populations in cluster dEs, while the central regions consist of young populations.  

\begin{figure*}
\centering
	\includegraphics[width=1\linewidth]{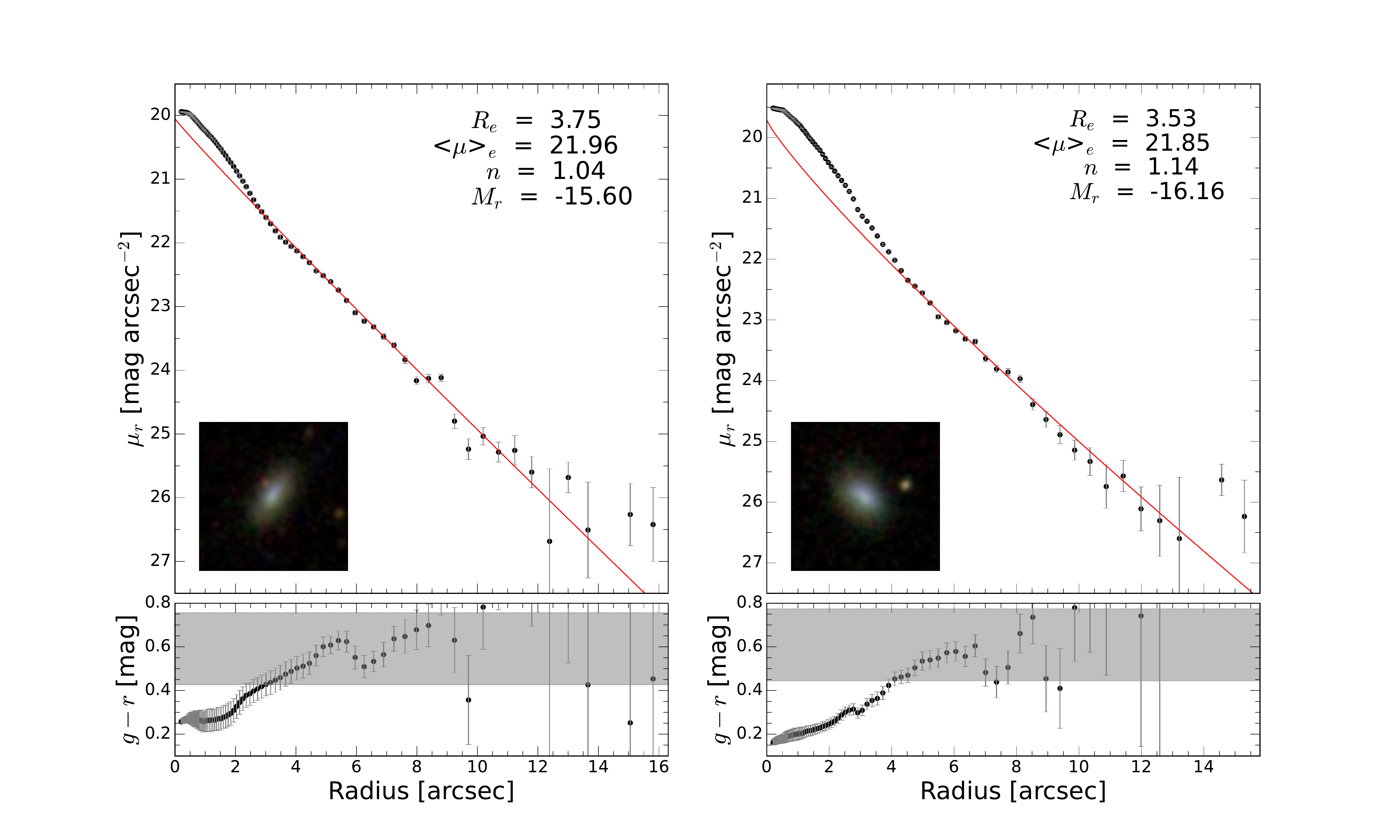} 
\caption{$Upper$: Examples of S\'{e}rsic fit (solid curve) to the SDSS $r$ band surface brightness profiles of isolated dE(bc)s. The inset shows the $g$, $r$, and $i$ bands combined color images of galaxies. $Lower$: Radial $g-r$ color profiles, where radius was calculated from the semimajor axis. The grey-shaded areas are the 2$\sigma$ range of the colors of typical dEs without blue cores at the respective magnitude derived from the CMR of the EVCC \citep{Kim2014}. In all panels, error bars denote the rms deviation for each radius.}    

    \label{fig:figure1}
\end{figure*}

The stellar masses ($M_{*}$) of isolated dE(bc)s were derived using a relation between the $g-i$ color and the stellar mass-to-light ratio based on $i$ band luminosity ($L_{i}$; \citealt{Bell2003}) assuming the initial mass function of \citet{Kroupa1993}:

\begin{equation} 
log(M_\star/M_\odot)=-0.152+0.518(g-i)+log(L_i/L_\odot). 
 \end{equation} 

Our selected isolated dE(bc)s are in the stellar mass range of 7.6 $< log M_{*} <$ 9.5 with a median value of 8.3.

\section{Results}
\subsection{Structural Parameters}
We compared the structural parameters of isolated dE(bc)s to those of dwarf galaxies in the Virgo cluster: dE(bc)s identified by \citet{Lisker2006a}, dEs, and  high surface brightness irregular galaxies (Irr(HSB)s) that are possible BCD candidates in morphological classification \citep[see][for details]{Kim2014}. Figure 2 presents the distributions of $M_{r}$ vs. $R_{e}$ and $<\mu>$$_{e}$ for the galaxies. Isolated dE(bc)s overlap with the Virgo dE(bc)s at bright magnitudes, suggesting that isolated dE(bc)s exhibit similar structural parameters to their Virgo counterparts. Note that dE(bc)s appear to be offset from the mean distribution of dEs and show a distribution comparable to that of Irr(HSB)s. 
At a fixed magnitude, the dE(bc)s are more compact (i.e., show smaller $R_{e}$) and have higher surface brightnesses than the majority of dEs, indicating the centrally concentrated SF activity in dE(bc)s \citep{Meyer2014,Chung2019}
.

\begin{figure*}
\centering
	\includegraphics[width=0.8\linewidth]{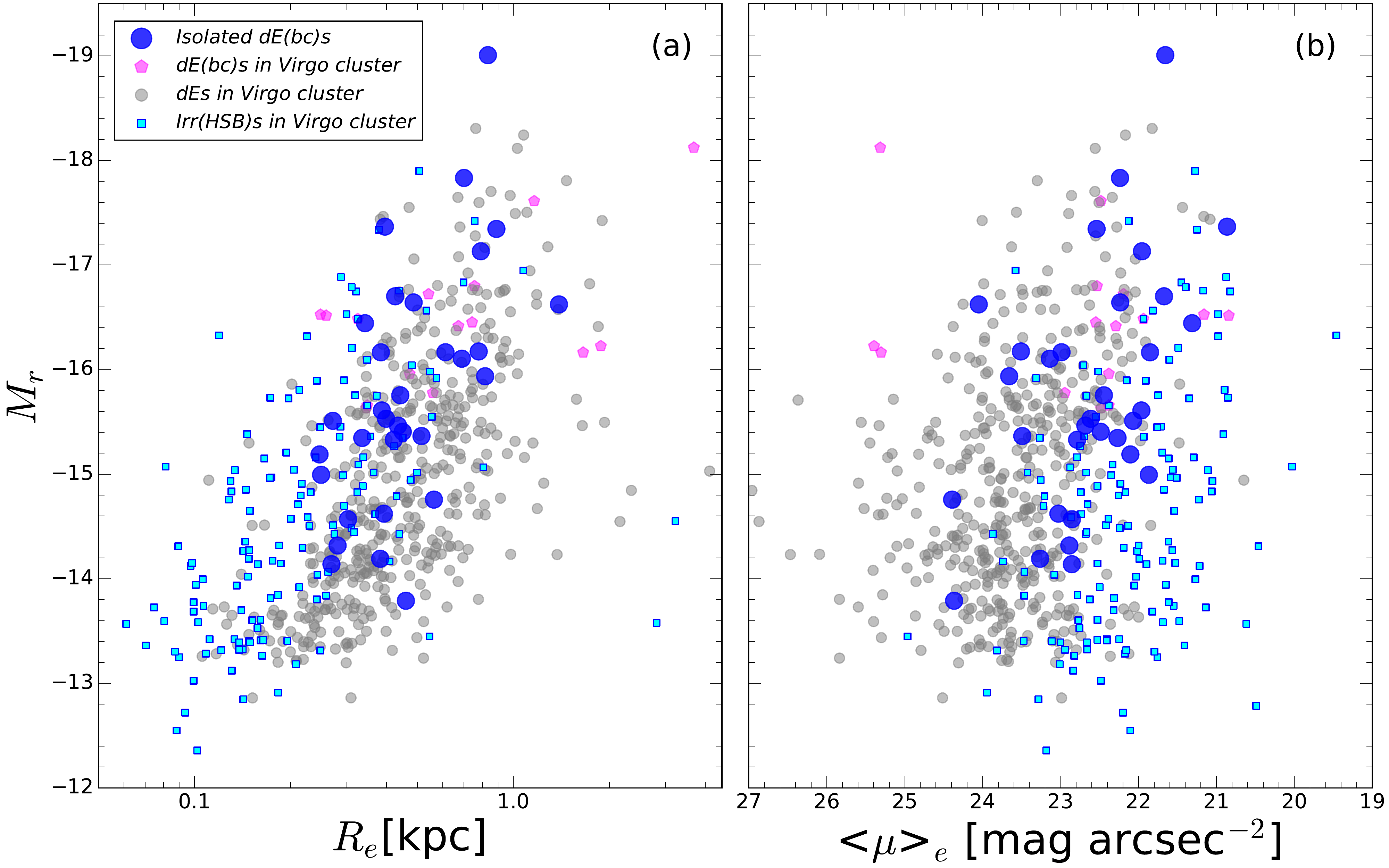} 
        \caption{Distribution of isolated  dE(bc)s (blue circles) in $(a)$ $M_{r}$ vs. $R_{e}$  and $(b)$ $M_{r}$ vs. $<\mu>$$_{e}$. As a comparison sample, we plot the dE(bc)s (pink pentagons), dEs (grey circles), and Irr(HSB)s (cyan squares) in the Virgo cluster. 
}
    \label{fig:figure2}
\end{figure*}

\subsection{Stellar Populations}
The near-$UV$ ($NUV$; 1750-2750\AA)$-$optical color is a particularly efficient tool for tracing recent SF since the $NUV$ flux originates from young massive stars with timescales of a few hundred Myr \citep{Kennicutt1998}. In Figure 3a, we compare the $NUV-r$ vs. $M_{r}$ color-magnitude relation (CMR) of isolated dE(bc)s to that of dE(bc)s in the Virgo cluster. For comparison, we also overplot dEs and Irr(HSB)s in the Virgo cluster drawn from the EVCC \citep{Kim2014}. The $NUV-r$ colors of the majority of isolated dE(bc)s and Virgo dE(bc)s are bluer than those of the red sequence of the Virgo cluster (dashed line) at a given magnitude. One notable feature is that isolated dE(bc)s have systematically bluer $NUV-r$ colors than those of the Virgo dE(bc)s at all magnitudes, and partly overlap with the distribution of Virgo Irr(HSB)s. 

We also show the $\Delta$($NUV-r$) vs. $g-r$ color gradient of galaxies in Figure 3b, where $\Delta$($NUV-r$) is the offset of the $NUV-r$ color from the red sequence of the Virgo cluster at a given magnitude in the CMR (i.e., color residual). A strong correlation is present in that dE(bc)s with bluer $NUV-r$ colors exhibit larger positive color gradients. Moreover, the color gradients of isolated dE(bc)s are systematically larger than those of Virgo dE(bc)s. Since the colors of the outer regions of most dE(bc)s are comparable to global colors of red sequence galaxies (see lower panels of Figure 1), the positive color gradients in the dE(bc)s are a result of the radial distribution of young stars. Therefore, more vigorous, centrally concentrated SF activities (i.e., a larger mass fraction of young stars) in isolated dE(bc)s would be responsible for their steeper color gradients and overall bluer $NUV-r$ colors compared to their Virgo counterparts.

\begin{figure*}
\centering
	\includegraphics[width=1.0\linewidth]{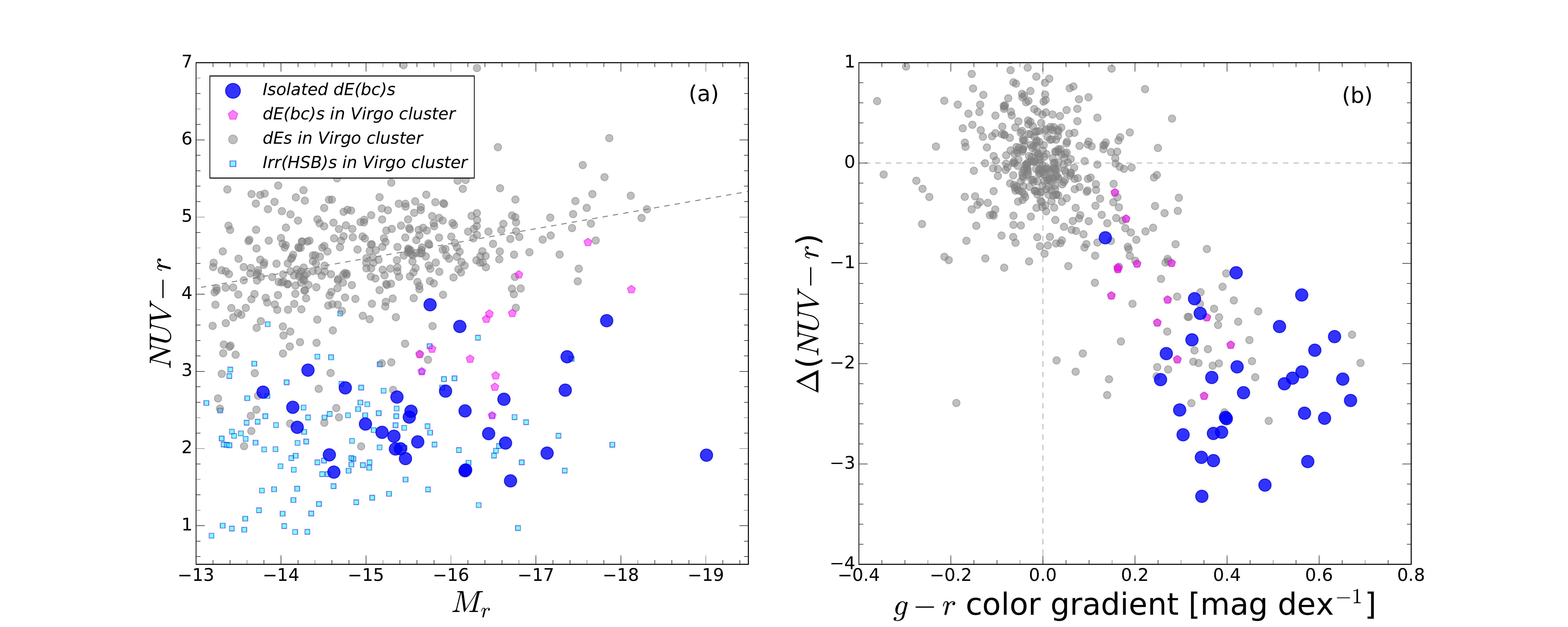} 
    \caption{$(a)$ $NUV-r$ vs. $M_{r}$ CMR of galaxies with various types. Symbols are the same as in Figure 2. The dashed line represents the linear least-square fit to the normal dEs (i.e., red sequence) in the Virgo cluster. $(b)$ $\Delta$($NUV-r$) vs. $g-r$ color gradient of galaxies, where $\Delta$($NUV-r$) is the deviation of the color of a galaxy from the red sequence of the Virgo cluster at a given magnitude in the CMR.}
    \label{fig:figure3}
\end{figure*}

The presence of H$\alpha$ emission clearly indicates the ongoing SF activity from most massive stars within the past a few tens of Myr  \citep{Kennicutt1998}. We extracted the equivalent width (EW) of the H$\alpha$ emission line of dE(bc)s from the NSA. All of isolated dE(bc)s exhibit a H$\alpha$ emission line with a median EW(H$\alpha$) value of 45 \AA. 
Consequently, the overall high strengths of $NUV$ flux and H$\alpha$ emission line of isolated dE(bc)s indicate that these galaxies have experienced active SF in their central regions.

We derived the H$\alpha$ SF rates (SFRs) of the isolated and Virgo dE(bc)s from the relation given by Kennicutt (1998) based on the H$\alpha$ emission line flux ($L_{H\alpha}$) taken from the NSA,
\begin{equation} 
SFR(M_\odot yr^{-1})=7.9\times10^{-42}L_{H\alpha}(ergs s^{-1}).
\end{equation} \

For correction of internal extinction to the $L_{H\alpha}$, we used the theoretical Balmer ratios of H$\alpha$/H$\beta$=2.86 for Case B emissivity with a temperature of 10,000 K and electron density of 100 cm$^{-3}$ \citep{Osterbrock2006}. The H$\alpha$ specific SFRs (sSFRs) of isolated dE(bc)s are in the range of  -13 $<$ log(sSFR) $<$ -10 with a median value of -10.5 that is about 1 dex larger than that of dE(bc)s in the Virgo cluster (i.e., -11.6). 

\subsection{Emission-Line Diagnostics}
In Figure 4, we plot the isolated and Virgo dE(bc)s in a Baldwin–Phillips–Terlevich (BPT) diagram \citep{Baldwin1981} using the emission line ratios [OIII]/H$\beta$ and [NII]/H$\alpha$ extracted from the NSA. We divide isolated dE(bc)s into two subsamples with the characteristics of BCDs and quiescent BCDs (QBCDs) based on the selection criteria of \citet{SanchezAlmeida2008}. 
The BCDs are characterized by their blue colors, compact morphology, low luminosities, large SFRs, and low gas metallicities (see Table 3 of \citealt{SanchezAlmeida2008} for details). The range of colors, magnitudes, and structural parameters (i.e., concentration indices) are also used for selecting QBCDs (see Table 1 of \citealt{SanchezAlmeida2008} for details).

For comparison, we also overplot BCDs (blue contours) and QBCDs (grey contours) in the nearby Universe selected by \citet{SanchezAlmeida2008} from the SDSS. We use the demarcation lines of the theoretical maximum starburst model (\citealt{Kewley2001}; dashed curve) and empirical star-forming limit (\citealt{Kauffmann2003}; solid curve) to separate star-forming galaxies from active galactic nuclei (AGNs) and transition galaxies (i.e., blend of star-forming and AGN features), respectively. All isolated dE(bc)s are classified as star-forming galaxies.

\begin{figure*}
\centering
	\includegraphics[width=0.5\linewidth]{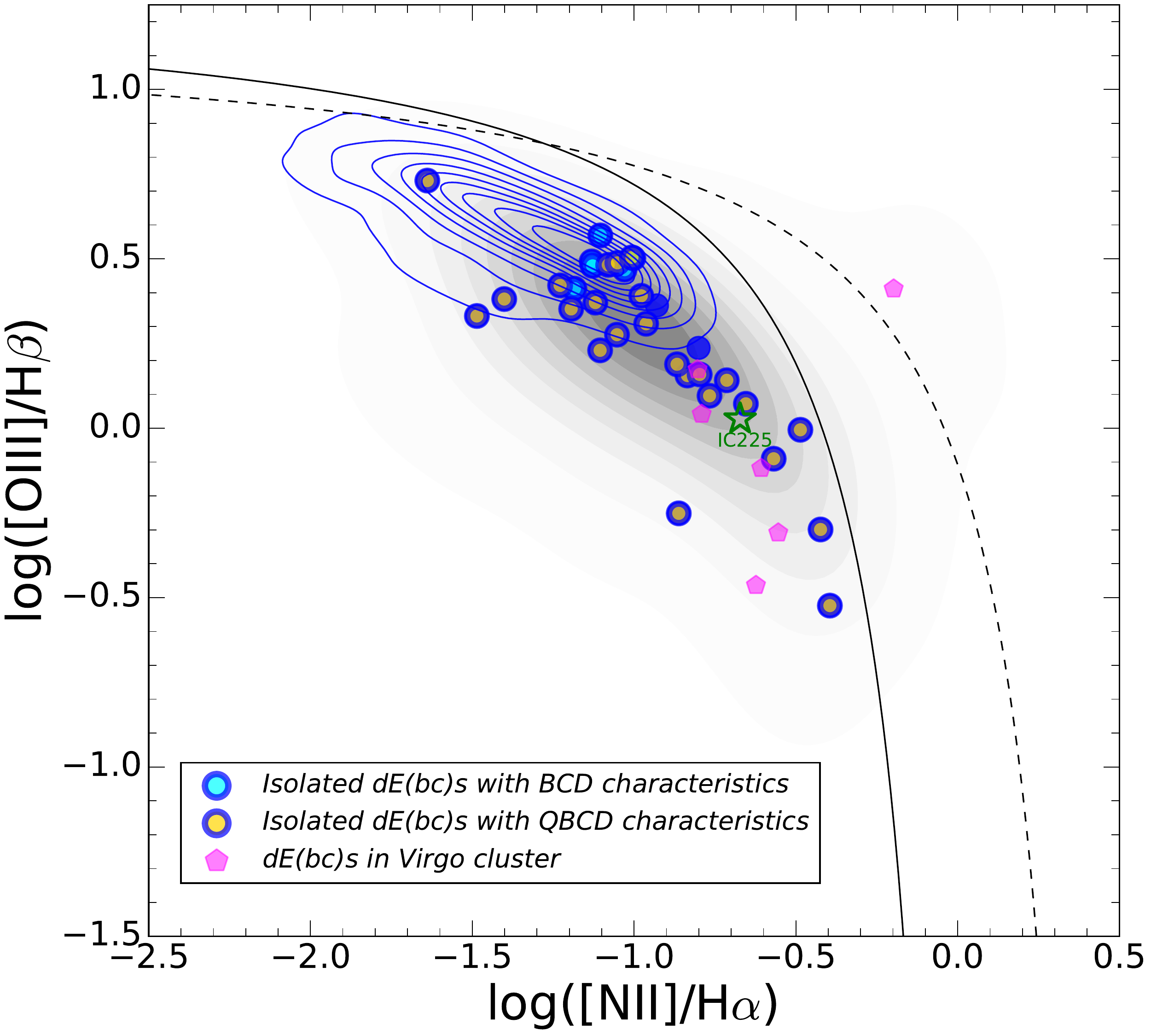} 
    \caption{BPT diagnostic diagram of isolated dE(bc)s with the characteristics of BCD (cyan circles) and QBCD (yellow circles). The two blue circles are isolated dE(bc)s that are not classified as BCD or QBCD. IC 225, an isolated dE(bc) discovered by \citet{Gu2006}, and Virgo dE(bc)s are respectively denoted by a green star and by pink pentagons. Blue and grey contours respectively denote the distributions of BCDs and QBCDs selected by \citet{SanchezAlmeida2008} from the SDSS.  
    Solid and dashed curves are the empirical star-forming limit \citep{Kauffmann2003} and theoretical maximum starburst model \citep{Kewley2001}, respectively.   
    }
    \label{fig:figure4}
\end{figure*}

\section{Discussion and Conclusions}
We have reported that most dEs discovered in isolated environments at z $<$ 0.01 exhibit blue cores in optical images and positive gradients in color profiles. From the $UV$ detection and optical emission lines, these isolated dE(bc)s show clear signs of more significant ongoing or recent central SF activity than their counterparts in the Virgo cluster. The SF of all isolated dE(bc)s must be transient from the exhaustion of their gas reservoirs unless gas inflow from external sources is permitted; for example, dE(bc)s are likely to evolve passively into red, quiescent galaxies after approximately 1 Gyr \citep[e.g.,][]{Ferreras2009,Suh2010}.  However, this seems to be at odds with the observational results at low redshift that quenched dwarf galaxies in isolation are statistically rare, but blue, star-forming dwarf galaxies, such as BCDs, are common in isolated environments \citep[e.g.,][]{Wang2009,Geha2012}. In this respect, isolated dE(bc)s could not be anticipated to evolve into quenched dEs.

Whereas the isolated BCDs exhibit global, intense starburst, we note that a large sample of isolated QBCDs without major starburst characteristics has also been identified \citep{SanchezAlmeida2008}. The BCDs and QBCDs show a close relationship in that they are morphologically and structurally similar \citep{SanchezAlmeida2008,Meyer2014}. Interestingly, many fractions of QBCDs exhibit morphologies with a hint of centrally concentrated SF that are similar to those of dE(bc)s. 

\citet{SanchezAlmeida2008} suggested that the recurrent evolutionary sequence of BCD and QBCD phases lasts for a few Hubble times
\citep[see also references in][]{SanchezAlmeida2008}.
In this sustainable two-phase scenario, BCDs have a short ($\sim$ 10 Myr) period of starburst, interleaved with long ($\sim$ 270 Myr) periods of quiescence (i.e., QBCD phase). The recursive inflow of gas may play a critical role in bridging between BCDs and QBCDs by regulating the SFR and, therefore, the global properties of the galaxies \citep[e.g.,][]{Stinson2007,Erb2008,SanchezAlmeida2008,SanchezAlmeida2014,Zhao2013,Saintonge2022}. The pristine metal-poor gas may exist in extensive, circumgalactic and intergalactic structures. If the gas falls on the QBCDs over long periods and the gas density exceeds the prerequisite threshold, the QBCDs can enter the BCD phase of a short starburst. The large gas outflows caused by the starburst of BCDs can lead to the long inter-burst QBCD phase with diminished SF. Later, the expelled gas falls back on the QBCDs again, creating a new starburst. Alternatively, the accretion of fresh intergalactic gas can also trigger SF \citep[e.g.,][]{Dekel2009}.

We find that most ($\sim$ 75 \%) isolated dE(bc)s are classified as QBCDs according to the selection criteria of \citet{SanchezAlmeida2008}. As shown in Figure 4, most isolated dE(bc)s show a distribution similar to that of QBCDs (grey contours) reported by \citet{SanchezAlmeida2008}. A wide distribution of QBCDs is consistent with a considerable variation in their observational properties \citep{SanchezAlmeida2008,SanchezAlmeida2009}. Given the heterogeneity of QBCDs, they might be at different evolutionary stages in the BCD-QBCD cycle. An overlap of the distribution of QBCDs with that of BCDs (blue contours) is also in line with these two galaxy populations being part of a repeated, continuous evolutionary sequence \citep[see also Figure 4 of][]{SanchezAlmeida2008}. Consequently, we suggest that isolated dE(bc)s might be QBCDs that move to or back from the BCD phase (i.e., pre- or post-BCDs).  

If the starburst of BCDs exhausts the gas reservoirs within a short timescale of 1 Gyr \citep[e.g.,][]{Lee2002}, this may not allow the cadence of BCD$-$QBCD phases to last during a galactic lifetime. However,  isolated starbursting dwarf galaxies have managed to retain gas over a long time scale (more than 10 Gyr) because stellar feedback is not very efficient in ejecting all of the gas from the galaxies \citep{MacLow1999,Ferrara2000, Tajiri2002,Valcke2008}. 
It is also possible that an external gas supply could be provided from the cosmic web \citep[see][and references therein]{SanchezAlmeida2014,Saintonge2022}.

In Figure 5, we plot the distribution of the atomic HI gas mass ($M$$_{\rm gas}$) fraction using data from ALFALFA \citep{Haynes2018}, defined as
$\mu$ = $M$$_{\rm gas}$/($M$$_{\rm gas}$ $+$ $M_{*}$), as a function of the stellar mass ($M_{*}$) of isolated dE(bc)s in comparison with that of dE(bc)s in the Virgo cluster. Additionally, we present the distribution of E-type BCDs \citep{Zhao2013} that have structural properties compatible with dEs \citep{Papaderos1996,GildePaz2005}.  At a given $M_{*}$, isolated dE(bc)s exhibit more significant gas mass fractions than the Virgo counterparts. Moreover, the distribution of isolated dE(bc)s is comparable to that of BCDs. Although the SF of isolated dE(bc)s is limited to the central region, this suggests that isolated dE(bc)s do have substantial gas reservoirs for subsequent active starburst BCD stage in the BCD$-$QBCD cycle.

\begin{figure*}
\centering
	\includegraphics[width=0.4\linewidth]{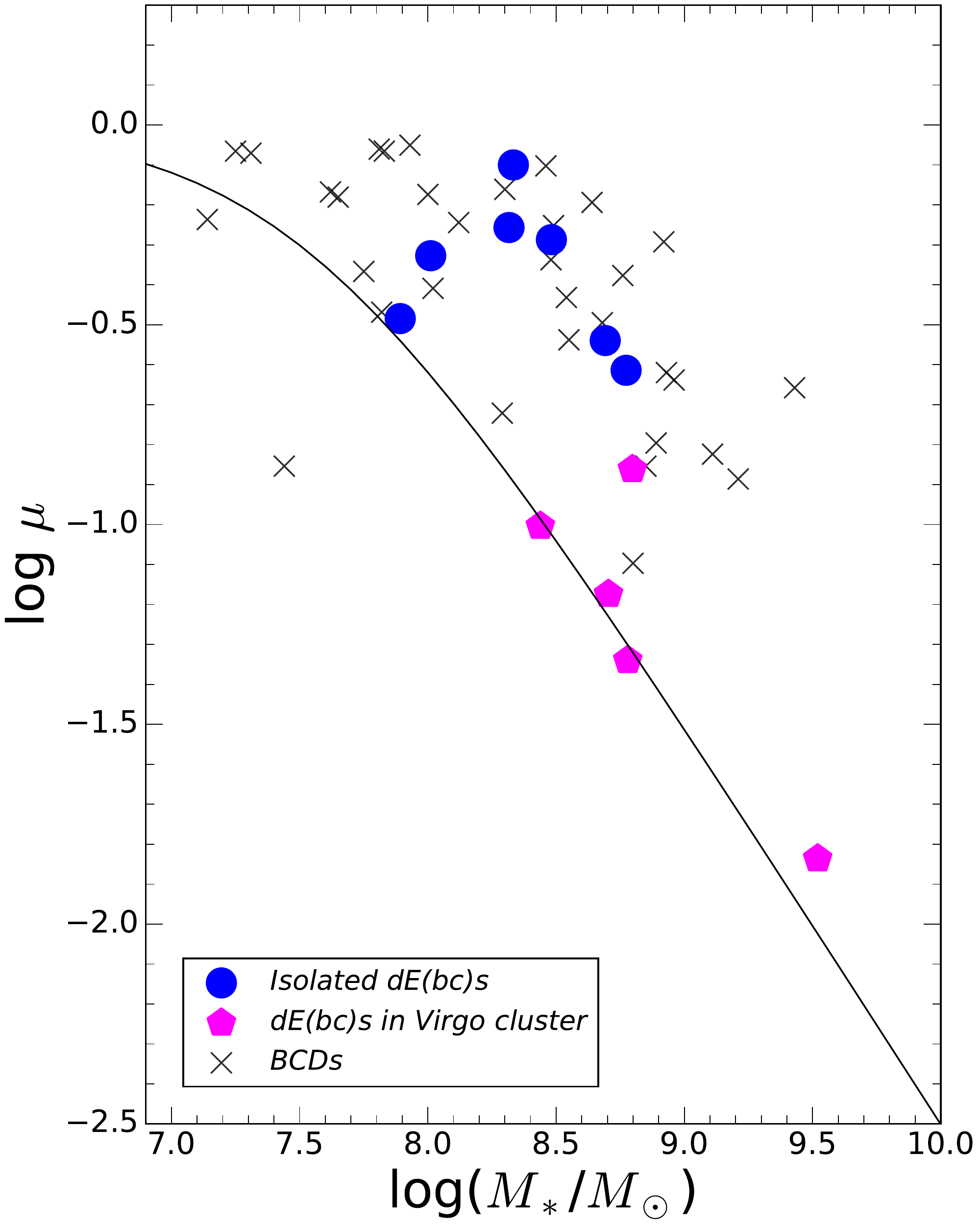} 
    \caption{HI gas mass fraction ($\mu$) vs. stellar mass distribution of isolated dE(bc)s (blue circles), Virgo dE(bc)s (pink pentagons), and E-type BCDs \citep[crosses;][]{Zhao2013}. The solid line denotes the limiting sensitivity of ALFALFA at the distance of the Virgo cluster. }
    \label{fig:figure5}
\end{figure*}

Extended gas distribution around a galaxy and external gas infall have been observed based on the HI mapping observations \citep[see][and references therein]{Sancisi2008, SanchezAlmeida2014,Saintonge2022}. For example, NGC 5253, a nearby BCD, is enclosed with a large-scale ($\sim$two to three times its optical size), massive ($\sim$10$^{8}$ M$_{\odot}$) HI structure and is experiencing the infall of HI cloud \citep{LopezSanchez2012}. If an isolated dE(bc) is surrounded by huge gas clouds extending to outside the virial radius of the galaxy, gas accretion can sustain SF as inferred from the scenario of the BCD$-$QBCD cycle over cosmic time.  Future wide-field, deep HI observations of mapping around isolated dE(bc)s will be crucial for understanding the gas distribution comprehensively.

It has been suggested that the transformation of late-type galaxies by environmental processes is the main channel for the formation of dE(bc)s in galaxy clusters \citep[][and references therein]{Chung2019}. However, the presence of isolated dE(bc)s implies that some dE(bc)s may not be the unique product of morphological transformation. Therefore, we suggest that at least some gas-bearing dE(bc)s found in clusters \citep[e.g.,][]{Lisker2006a,Hallenbeck2012,DeRijcke2013} might be initially isolated dE(bc)s that subsequently fell into the clusters. Once these galaxies enter the cluster environment that is not conducive to retaining cold gas, they will rapidly approach the end of their SF and eventually become passive dEs. Finally, we propose that dE(bc)s are a mixture of objects formed by two main channels (i.e., nature or nurture) depending on the environment (i.e., isolated or high-density).

\begin{acknowledgments}
We are grateful to the anonymous referee for helpful comments and suggestions that improved the clarity and quality of this paper. This work was supported by the National Research Foundation of Korea through grants NRF-2022R1A2C1007721 (S.C.R.), NRF-2019R1I1A1A01061237 \& NRF-2022R1C1C2005539 (S.K.), NRF-2018R1A6A3A01013232 \& NRF-2022R1F1A1072874 (J.C.), and NRF-2022R1I1A1A01054555 (Y.L).
\end{acknowledgments}


\bibliography{sample631}{}
\bibliographystyle{aasjournal}



\end{document}